\documentclass[overfull,debug]{epl}
\usepackage{epsfig,amsmath,amssymb}

\title{Phase behaviour of a symmetrical binary fluid mixture in a field}
\shorttitle{Symmetrical mixtures in a field}
\author{J\"{u}rgen K\"{o}finger \inst{1}, Gerhard Kahl, \inst{2} and Nigel B. Wilding \inst{3}}
\shortauthor{K\"{o}finger {\em et al}}
\institute{
\inst{1} Faculty of Physics, University of Vienna, Boltzmanngasse 5 , 1090 Vienna, Austria\\
\inst{2} Institut f\"ur Theoretische Physik and Center for Computational
Materials Science -  TU Wien, Wiedner Hauptstra{\ss}e 8-10, A-1040 Wien, Austria\\
\inst{3} Department of Physics, University of Bath, Bath BA2 4LP, United Kingdom
}

\pacs{05.70.jk}{First pacs description}
\pacs{64.70.Fx}{Second pacs description}
\pacs{64.70.Ja}{Third pacs description}

\begin{document}

\maketitle

\begin{abstract}  

Integral equation theory calculations within the mean spherical
approximation (MSA) and grand canonical Monte Carlo (MC) simulations
are employed to study the phase behaviour of a symmetrical binary
fluid mixture in the presence of a field arising from unequal chemical
potentials of the two particle species.  Attention is focused on the
case for which, in the absence of a field, the phase diagram exhibits
a first order liquid-liquid transition in addition to the liquid-vapor
transition. We find that in the presence of a field, two possible
subtypes of phase behaviour can occur, these being distinguished by
the relationship between the critical lines in the full phase diagram
of temperature, density, and concentration. We present the detailed
form of the respective phase diagrams as calculated from MSA and
compare with results from the MC simulations, finding good overall
agreement.

\end{abstract}


\section{Introduction}

The phase behaviour of binary symmetrical liquid mixtures has attracted
considerable interest over many years
\cite{Pan88,Rec93,Gre94,deM95,Wil97,Cac98,Wil98,Ant02,Wil03,Schoen03,Sch03,Pin03,Sch04,Sch05,Sch05a}.
In such a system the interaction between particles of the same species
(labeled `1' and `2') is equal, i.e., $\Phi_{11}(r) = \Phi_{22}(r)$, but
that between unlike species is modified by a factor $\delta$, so that
$\Phi_{12}(r) = \delta \Phi_{11}(r)$. This model serves as a starting
point for studying the phase behaviour of more realistic binary fluids.
Furthermore it maps onto an Ising spin fluid and therefore
represents a prototype one-component fluid in which the particles carry
an internal degree of freedom. 

The topology of the phase diagram is governed by the parameter
$\delta$.  We shall only consider the case of $\delta < 1$, for which
there is a demixing transition.  In the equimolar case (i.e., when
both species have equal concentrations), one finds four different
topologies of phase diagram \cite{Wil98,Sch05a}.  These are
distinguished by the relationship between a critical line of demixing
transitions and a first order liquid-vapour (LV) transition.
Interestingly it transpires that the same four archetypes of phase
diagram occur in considerably more complex and realistic models such
as Heisenberg and Stockmayer fluids (see \cite{Tav95,Wei97,Gro97} and
references therein).  Thus the simple symmetrical mixture provides a
useful test bed for elucidating the generic aspects of phase behaviour
without the need to tackle these more complicated systems explicitly.

Depending on how the critical line of the demixing transition (the
so-called $\lambda$-line) intersects the LV coexistence line in the
density-temperature plane, the four types of phase diagram can briefly
be described as follows (in order of decreasing $\delta$): in the
first type, the $\lambda$-line intersects the LV coexistence line from
the liquid side at temperatures well below the LV critical
temperature, forming a critical end point (CEP); in the second type,
the $\lambda$-line approaches LV coexistence (from the liquid side)
slightly below the LV critical temperature, terminating in a separate
tricritical point, which itself marks the end of a first order
transition between mixed and demixed liquid phases; in the third type
the $\lambda$-line intersects the LV critical point directly, the two
combining to form a tricritical point.  Finally in the fourth type the
$\lambda$-line intersects the low density branch of the LV coexistence
curve, forming a CEP. We point out that these topologies are special
cases of the more general classification scheme for the phase behaviour of
binary fluid mixtures proposed by van Koynenburg and Scott
\cite{Scott}; the correspondence with the latter has been described in
ref. \cite{Sch05a}.

These four types of phase diagram have been studied in detail for
several symmetrical systems in the case of equimolar concentrations of
the two species. This corresponds to a vanishing chemical potential
difference field, i.e. $\Delta \mu = \mu_1 - \mu_2 = 0$. We shall
refer to this scenario as the field-free case. For $\Delta \mu \ne 0$
we are aware of only two studies \cite{Pin03,Ant02} for continuum
models. This is probably related to the fact that the numerical
solution of the coexistence equations is cumbersome and that the
topology of the {\it full} phase diagram becomes rather intricate in
the absence of the equimolar symmetry.

It is, in particular, the remarkable study of Pini {\it et al.}
\cite{Pin03} that has provided a first impression of the unexpected
richness of the {\it full} phase diagram of a binary symmetrical
mixture. Focusing on the topology of phase diagram in which, for $\Delta
\mu =0$, a first order liquid-liquid transition occurs in addition to a
LV transition, the authors report for the case $\Delta \mu \ne 0$ the
occurrence of double critical and tricritical points, and observe a
region of a homogeneous, mixed fluid. This investigation was based on
both a simple mean-field (MF) approach and Hierarchical Reference Theory
(HRT) calculations, the latter being a highly sophisticated liquid state
theory that remains accurate even near critical points (see \cite{Pin03}
and references therein). However, as a consequence of the high numerical
costs of such calculations, a complete elucidation of the phase
behaviour was out of reach; the study instead being limited to a few
isothermal cuts at selected $\delta$ values.

In the present work we have re-considered exactly the same system as
in \cite{Pin03}, namely a binary symmetrical mixture whose particles
interact via a hard-core Yukawa (HCY) potential in an external
field. However, for reasons of computational efficiency we have
studied its phase behavior by means of the mean spherical
approximation (MSA), taking advantage of the availability of the
analytical solution of the MSA for HCY mixtures
\cite{Hoy77,Hoy78,Arr87,Arr91}. Doing so allows us to explore the {\it
full} phase diagram for $\Delta \mu \ne 0$ in a systematic fashion,
something which would be far more arduous using HRT. We complement our
MSA study by grand-canonical Monte Carlo (MC) simulations. Both the
theory and the simulations confirm the existence of the two subtypes
of phase behaviour in the presence of a field, which we label $\alpha$
and $\beta$, and are distinguished by the trajectories of the critical
lines in the phase diagram. In the classification scheme of phase
diagrams of binary mixtures \cite{Scott} subtype $\alpha$ corresponds
to type `sym. III-A$^\star$', and subtype $\beta$ to type
`sym. II-A$^\star$'. We present these phase diagrams in the three
dimensional (3d) space of temperature $T$, number density $\varrho =
(N_1+N_2)/V$, and concentration $c=N_1/N$, thus permitting a deeper
insight into the respective topologies than has hitherto been
possible. The new perspectives thus gained, allow further elucidation
of the nature of the initial findings presented in \cite{Pin03} and a
deeper appreciation of their wider context. Additionally, we uncover
new phenomena in this context (such as triple lines) that have not, to
date, been reported. Our comparison of theory with the results of
computer simulations shows that the latter are competitive with
theoretical approaches in describing even complex phase phenomena on a
{\it quantitative} level.

\section{System and methodology} 

We consider a binary symmetrical mixture where the particles interact via the HCY potential

\begin{equation}                                  \label{phi_tot}
\Phi_{ij}(r) = \left\{ \begin{array} {l@{~~~~~~~~}l}
                             \infty  & r \le \sigma \\
                             -\epsilon_{ij} \sigma \frac{1}{r}
                             \exp[-z (r - \sigma)]    & r > \sigma \\
                             \end{array}
                             \right. .
\end{equation}
Here $\epsilon_{ij}$ are the contact values of the potentials, $\sigma$
is the hard-core diameter, and $z$ is the inverse screening length, assigned the value $z=1.8/\sigma$ in this work. 
Standard reduced units will be used throughout the paper.

The solution of the Ornstein-Zernike equations \cite{Han86} along with
the MSA closure relation, i.e.,

\begin{equation} c_{ij}(r) = - \beta \Phi_{ij}(r) ~~~ r > \sigma ~~~~~
{\rm and} ~~~~~  g_{ij}(r) = 0 ~~~ r \le \sigma 
\end{equation} 
can be carried out for this particular system to a large extent
analytically \cite{Hoy77,Hoy78}.  Here $c_{ij}(r)$ and $g_{ij}(r)$ are
the direct correlation functions and the pair distribution functions
respectively, while $\beta=1/k_\mathrm{B}T$ is the inverse
temperature.  We have used the formalism presented by Arrieta {\it et
al.}  \cite{Arr87,Arr91} for the MSA solution for multi-component HCY
systems. These papers contain expressions for thermodynamic
properties, notably the pressure and the chemical potentials which are
required to determine phase coexistence, as well as details of the
numerical solution of the set of coupled, non-linear equations that
fix the unknown parameters of the correlation functions. All our MSA
calculations have been carried out using {\tt
MATHEMATICA}\texttrademark \cite{Math}.

Grand canonical MC simulations were performed in a manner similar to
that employed for a previous study \cite{Wil98} of the zero field case
$\Delta \mu=0$.  Simulations were conducted at constant $T$, $\mu_1$,
and $\Delta\mu$ and measured the joint distribution of the number
density, the concentration, and the internal energy, which was
accumulated in the form of a list.  Phase boundaries were traced in
the space of these variables by applying the equal peak weight
criterion to the order parameter distribution \cite{Borgs}.  This
procedure was aided by multicanonical preweighting and multiple
histogram reweighting \cite{Berg, Ferrenberg}.  Critical points were
located using a crude version of the finite-size scaling techniques
described in ref.~\cite{Wil95}. This involves matching the order
parameter distribution function to a known universal scaling form
appropriate to the finite-size limit.

Further details concerning all methods used will appear in a forthcoming report \cite{Koe06}.

\section{Results} 

\begin{figure}[t]
\epsfig{angle=0, file=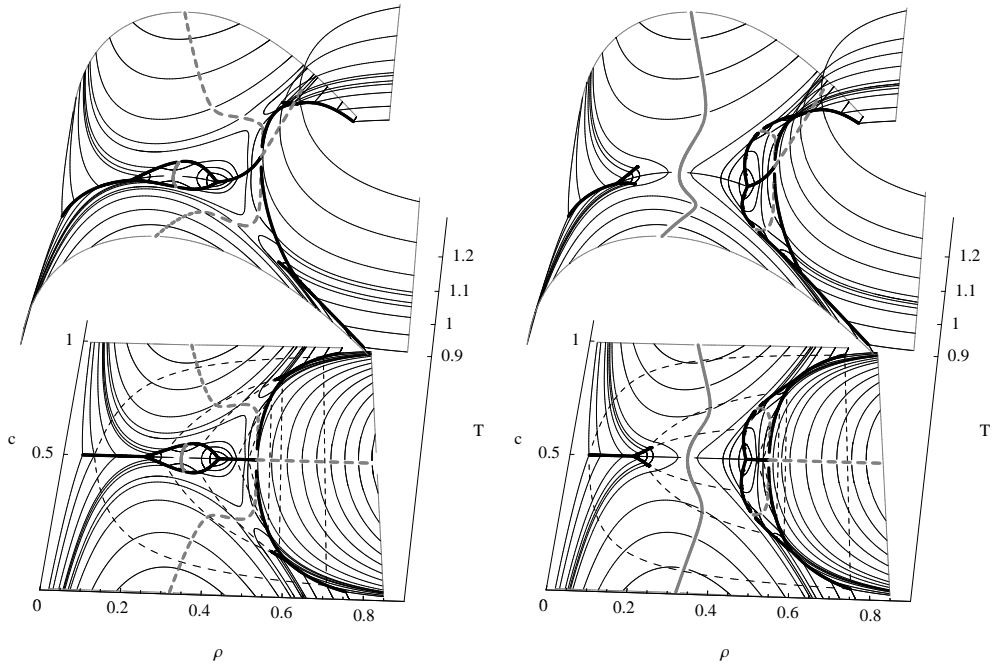, width=14.0cm, clip}
\caption{MSA phase diagram of the binary symmetrical mixture
considered in the study in $(T, \varrho, c)$-space and its projection
onto the ($c,\varrho$)-plane.  Left: $\delta=0.67$ - (a) and right:
$\delta=0.69$ - (b).  Symbols: thin full lines - isothermal coexistence lines, dashed
thin lines - tie lines, grey full thick lines - critical lines passing
through the LV critical point of field-free case, dashed thick lines -
critical lines passing through the tricritical point of the field-free
case, black thick lines - triple lines.  }
\label{MSA_both}
\end{figure}

We first present MSA results for the two subtypes of phase diagrams
($\alpha$ and $\beta$) in $(T, \varrho, c)$-space and their
projections onto the $(\varrho, c)$-plane in
Fig.~\ref{MSA_both}. These provide an overview of the principal
features of the respective topologies. Because of the symmetry of the
underlying model, all phase diagrams are symmetric with respect to the
plane $c=1/2$.

The subtype $\alpha$ is displayed in Fig.~\ref{MSA_both}(a) where,
motivated by the results of Pini {\it et al.} \cite{Pin03}, we have
chosen $\delta = 0.67$.  We observe four distinct coexistence surfaces
separated by triple lines (to be discussed below) and each exhibiting
a critical line.  The first one, $\mathcal{S}_1^\alpha$, is
symmetrical with respect to the ($c=1/2$)-plane and represents the
demixing transition which is well-known from the field-free case; its
critical line is the $\lambda$-line. On this surface two symmetrical
high-density phases ($\varrho \gtrsim 0.5$), specified by $(\varrho,
c)$ and $(\varrho, 1-c)$, are in coexistence. Next, two further
(symmetrically related) coexistence surfaces, $\mathcal{S}_2^\alpha$
and $\mathcal{S}_3^\alpha$, are encountered for low and intermediate
densities $(\varrho \lesssim 0.6)$; they describe neither pure
demixing transitions, nor pure LV transitions, hence the respective
order parameter is a linear combination of the density and
concentration differences of the two coexisting phases. Coming from
high temperatures the $\lambda$-line bifurcates at the tricritical
point into two critical lines (belonging to $\mathcal{S}_2^\alpha$ and
$\mathcal{S}_3^\alpha$) which pass through minima and head toward the critical points of the
pure phases.  
The fourth coexistence surface, $\mathcal{S}_4^\alpha$,
which is located in between $\mathcal{S}_2^\alpha$ and
$\mathcal{S}_3^\alpha$, is predominantly LV-like in character and
contains a critical line of LV transitions which passes through the LV
critical point of the field-free mixture.  The surface as a whole is
delimited by the intermediate-density branches of a triple line which
form a lens-shaped `loop'. Tie-lines starting at this enclosing triple line connect a vapour and a liquid
phase of approximately equal concentrations with a liquid phase of
higher density. The latter states form the high-density branch of the
triple line located in the `valley(s)' formed by
$\mathcal{S}_2^\alpha$ (or $\mathcal{S}_3^\alpha$) and
$\mathcal{S}_1^\alpha$. In Fig.~\ref{MSA_both}(a) selected tie lines
are shown. In the field-free phase diagram four phases coexist, two of
which, being located on the high density $\mathcal{S}_1^\alpha$
surface, are symmetrically related.  The other two, at intermediate
densities, are the end points of the lens-shaped `loop' mentioned
above.  As the critical line of $\mathcal{S}_4^\alpha$ approaches the
bounding triple line, the coexisting vapour and liquid phases become
critical. Since these are simultaneously in equilibrium with a
non-critical phase (the so-called spectator phase) located at the end point of the high-density
branch of the triple line, this point is a CEP. Finally, we point out
that in agreement with \cite{Pin03} we also observe a region of a
homogeneous, mixed fluid at intermediate densities which is bounded by
the four coexistence surfaces.

In subtype $\beta$, which we have found to occur at $\delta = 0.69$, we
observe an entirely different topology of phase diagram, as is evident
in Fig.~\ref{MSA_both}(b). Again, four coexistence surfaces with
accompanying critical lines can be identified. At high densities we find
a symmetric demixing surface $\mathcal{S}_1^\beta$, associated with the
$\lambda$-line.  As the temperature is decreased, the $\lambda$-line
bifurcates at the tricritical point into two critical lines which
traverse the (symmetrically related) surfaces $\mathcal{S}_2^\beta$ and
$\mathcal{S}_3^\beta$.  These critical lines pass through minima and
terminate in CEPs located at the high density branch of a triple line
being the intersection of $\mathcal{S}_2^\beta$ and
$\mathcal{S}_3^\beta$ with a further coexistence surface
$\mathcal{S}_4^\beta$.  At these CEPs two high-density phases become
critical while they coexist with the spectator phase, located at the end
point of the low-density branch of the triple line.  A fourth critical
line on $\mathcal{S}_4^\beta$ connects the critical points of the
respective pure components; it passes through the equimolar LV critical
point and is now completely detached from the $\lambda$-line.  The
triple lines also show a distinctively different behaviour than their
equivalents in subtype $\alpha$: now one low-density phase (with
$c\simeq 1/2$) is connected via tie lines to two high-density phases;
representative tie lines are shown in Fig. \ref{MSA_both}(b).  From each
of the four points in the field-free phase diagram which coexist at the
triple point, a pair of triple lines eminates on application of an
external field. In contrast to type $\alpha$, the high and the
intermediate triple lines merge to form two symmetrically related
`loops', while the low density branches terminate as the spectator phase
of the CEP.  Again, for $\varrho \sim 0.5$ a region of a homogeneous,
mixed fluid is encountered, which is enclosed by the three coexistence
surfaces (see also \cite{Pin03}).

Of course it is of particular interest how the transition from one
subtype to the other takes place. We start from subtype $\alpha$ and
increase the parameter $\delta$. Then gradually the high- and the
intermediate-density branches of the triple line lengthen; they merge
when the CEP located on the intermediate-density branch meets the
end point of the high-density branch, thereby forming a tricritical
point. Within the MSA this occurs for $\tilde \delta_{\rm MSA} =
0.678(0)$. Upon further increase of $\delta$, a high-density branch
(with a CEP) and a low-density branch of the triple line detach, thus
resulting in the topology of the $\beta$-subtype. Concommitant with
this metamorphosis of the triple lines, we observe a related
development in the short section of the critical line that passes
through the LV critical point in subtype $\alpha$: with increasing
$\delta$, this line lengthens in both directions until it meets (at
the crossover between the subtypes) the critical lines of the surfaces
$\mathcal{S}_3^\alpha$ and $\mathcal{S}_4^\alpha$.  The critical lines
together form a loop as is nicely depicted in Fig. 1(c) of
\cite{Pin03} where, in the MF scenario, the transition occurs at
$\tilde \delta_{\rm MF} = 0.65338$. As $\delta$ is further increased,
a new critical line forms connecting the critical points of the pure
phases, passing through the equimolar LV critical point so that the
CEPs remain with the newly formed triple lines at higher densities.

With this picture in mind, we can now return to the study presented in
\cite{Pin03} and re-consider their results in a broader context.  Several
interesting features were discovered in that contribution and  depicted
via isothermal cuts through the phase diagram. 
However, we feel that a more comprehensive picture only emerges  
once the phase behaviour is depicted in the full $(T, \varrho,
c)$-representation of the phase diagram as presented here. In
particular we point out the following: (i) from our 3d representation we
obtain clear confirmation of the existence of two distinctly different
subtypes; (ii) we find previously unreported triple lines; (iii) our 3d
plots explicitly show how the reported `double critical points' in
ref.~\cite{Pin03} are related to local minima in critical lines.

Turning now to the results of our grand canonical MC simulations,
these involved tackling interesting features not previously considered
in the simulation literature, such as critical points for which the
order parameter is neither purely the density nor purely the
concentration, but a linear combination of the two.  In a manner
similar to \cite{Wil98} we accumulated the field-free ($\Delta \mu
=0$) phase diagram in the ($\mu_1, T$)-plane.  By means of histogram
reweighting we then extrapolated to small but finite $\Delta \mu$,
thus obtaining an estimate of the corresponding phase diagram.  Guided
by this prediction, a new set of simulations were then performed at
near coexistence state points for this value of $\Delta \mu$, the
results of which were extrapolated to yet larger $\Delta \mu$.  In
this manner we were able to track the phase behaviour as a function of
$\Delta \mu$, thereby permitting the study of a large range of
concentrations.  By accumulating separately contributions to the
energy from like and unlike particle interactions, we were also able
to perform histogram extrapolation with respect to $\delta$.  This was
useful in helping to find the regions of $\delta$ relevant to the two
subtypes.

\begin{figure}[!htb]
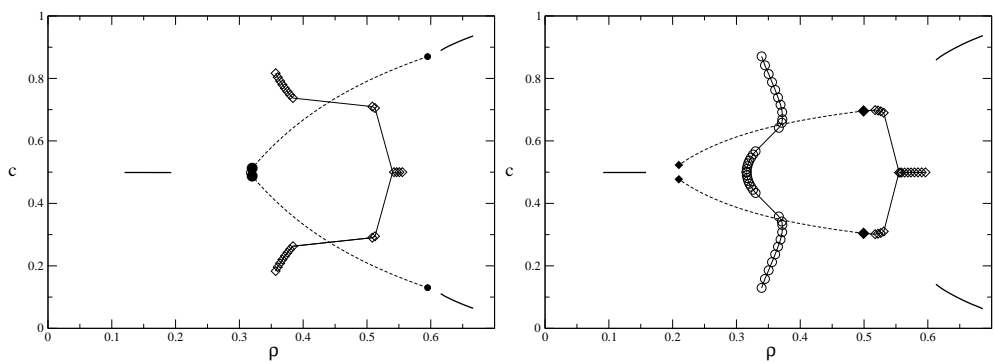

\epsfig{angle=0, file=./epl_d66_crho_final.eps, width=6.5cm, clip}
\epsfig{angle=0, file=./epl_d68_crho_final.eps, width=6.5cm, clip}
\caption{(a) Phase diagram as estimated by MC simulation and projected
onto the ($c,\varrho$)-plane for \mbox{$\delta=0.66$}. Thick solid lines
are field-free coexistence lines. 
Large circles mark critical points belonging to the critical line going through the field-free LV critical point, whereas diamonds represent critical points belonging to critical lines meeting in the tricritical point. 
Large filled symbols denote CEPs and small filled symbols represent the corresponding spectator phases. 
Dashed lines are tie lines and thin solid lines connect critical points belonging to the same critical line as a guide to the eye. 
\mbox{(b) The corresponding results for $\delta=0.68$.}}
\label{MC_both}
\end{figure}

Owing to the high dimensionality of the off-plane phase behaviour we
were not able to obtain the statistics necessary to construct the
entire 3d phase diagram in the same detail as done for
MSA. Nevertheless projections onto the $(\varrho, c)$-plane, as shown
in Fig.~\ref{MC_both}, clearly confirm the general picture seen in the
corresponding projections of the MSA data (Fig.~\ref{MSA_both}) and
outlined above. Specifically, for $\delta=0.66$, the critical line
eminating from the field-free LV critical point terminates at CEPs.
The critical lines eminating from the tricritical point link up with
that coming from the LV critical points of the pure phases. In
contrast, for $\delta=0.68$, the LV critical point of the pure phases
join smoothly to that of the equimolar mixture, while the critical
line starting from the tricritical point terminate at CEPs.  As
regards the range of $\delta$ in which the two subtypes of phase
behaviour are visible, the simulation results are in semi-quantitative
agreement with the MSA calculations, the value of $\delta$ separating
the two subtypes differs from that found in MSA by just $0.01$.

Further details of the MC and MSA results will be presented in a forthcoming
contribution\cite{Koe06}. 

\section{Conclusions}

Although the model system we have studied is a rather simple model of
a binary fluid, it nevertheless captures all the relevant features of
the phase behaviour of more complicated systems whose particles are
endowed with an internal degree of freedom, such as spin fluids. It
can therefore be viewed as a prototype model, the elucidation of which
represents the first step in a progression to other, more realistic
systems.

In this work we have shown that MSA provides a complete picture of the
two subtypes of phase behaviour for symmetrical mixtures in a chemical
potential difference field. The phase behaviour is considerably richer
in both variety and character than one would have expected on the
basis of the field-free case. Our results are in good agreement with
the less comprehensive MF and HRT study of these two subtypes in
ref.~\cite{Pin03}.  Although in principle more accurate than MSA, HRT
is very laborious to implement and computationally expensive, and
still does not produce results in fully quantitative agreement with
simulation, as has been observed in studies of the field-free case
\cite{Wil03}.

Our results are confirmed by MC simulations, showing that the MSA
provides a qualitatively correct description of this system.  Indeed
our study has demonstrated that computer simulations are competitive
with theory in providing (within a reasonable amount of time)
information on intricate phase diagrams exhibiting complex topologies
of critical lines. Given that the various commonly used theoretical
approaches (HRT, MSA,$\ldots$) do not always agree, MC thus provides
an invaluable benchmark with which to compare.  Comparison with
simulation is thus generally desirable.

\acknowledgments JK and GK acknowledge financial support by the
\"Osterreichische Forschungsfond (FWF) under Project Nos. P15758-N08
and P17823-N08, the Hochschuljubil\"aumsstiftung der Stadt Wien under
Project No. 1080/2002, and the Au{\ss}eninstitut der TU
Wien. Additional financial support was provided by the Anglo-Austrian
ARC Programme of the British Council. The authors would like to thank
Davide Pini (Milan) for useful discussions.

%


\end{document}